\documentclass[12pt,preprint]{aastex}

\slugcomment{Accepted by The Astronomical Journal}

\shorttitle{Radio continuum and water masers in CB 54}
\shortauthors{de Gregorio-Monsalvo et al.}

\begin{document}


\title{Radio continuum emission and water masers towards CB 54}


\author{Itziar de Gregorio-Monsalvo\altaffilmark{1}, Jos\'e  F. G\'omez\altaffilmark{2}, Guillem Anglada\altaffilmark{2}, Jos\'e M. Torrelles\altaffilmark{3}, Thomas B. H. Kuiper\altaffilmark{4}, Olga Su\'arez\altaffilmark{2,5}, Nimesh A. Patel\altaffilmark{6} } 


\altaffiltext{1}{European Southern Observatory, Alonso de C\'ordova 3107, Vitacura, Casilla 19001, Santiago 19, Chile}

\altaffiltext{2}{Instituto de Astrof\'{\i}sica de Andaluc\'{\i}a (CSIC), Apartado 3004, E-18080 Granada, Spain}

\altaffiltext{3}{Instituto de Ciencias del Espacio (CSIC) and Institut d'Estudis Espacials de Catalunya, C/Gran Capit\`a 2-4, E-08034 Barcelona, Spain.}

\altaffiltext{4}{Jet Propulsion Laboratory, California Institute of Technology, USA}

\altaffiltext{5}{UMR 6525 H. Fizeau, Université de Nice Sophia Antipolis, CNRS, OCA. Parc Valrose, F-06108 Nice Cedex 2, France}

\altaffiltext{6}{Harvard-Smithsonian Center for Astrophysics, 60 Garden Street, Cambridge, MA 02138, USA}


\begin{abstract}
We present high angular resolution observations of water masers at 1.3~cm and radio continuum emission at 1.3, 3.6 and 6~cm towards the Bok globule CB 54 using the Very Large Array. 
At 1.3~cm, with subarsecond angular resolution, we detect a radio continuum compact source located to the south-west of the globule and spatially coincident with a mid-infrared embedded object (MIR-b). The spectral index derived between 6 and 1.3 cm ($\alpha$=0.3$\pm$0.4) is flat, consistent with optically thin free-free emission from ionized gas. We propose the shock-ionization scenario as a viable mechanism to produce the radio continuum emission observed at cm frequencies. 
Water masers are detected at two different positions separated by 2.3$\arcsec$, and coincide spatially with two mid-infrared sources: MIR-b and MIR-c. The association of these mid-IR sources with water masers confirms that they are likely protostars undergoing mass-loss, and they are the best candidate as driving sources of the molecular outflows in the region.

\end{abstract}





\section{Introduction}

The Bok globule \object{CB 54} is an active star forming region associated with the Vela OB1 cloud complex and located at 1.5 kpc \citep{Lau97}. Due to their simplicity and relatively small size ($<$10$\arcmin$), Bok globules offer an unique environment to study the outcome of the star formation processes in a relatively idealistic and isolated way \citep{Bok47,Cle88}. 

CB 54 shows different signposts of multiple star formation. The region contains several molecular outflows. There is a main bipolar CO outflow oriented in the northeast-southwest direction and centered near the IRAS point-like source \object{IRAS 07020-1618} (also named CB 54 YC1; \citealt{Yun94b}). In addition, several  H$_{2} $[$v$=1-0 S(1)] line emission knots detected by \citet{Kha03} suggest the presence of a second outflow in the east-west direction, which indicates the existence of different driving protostars. In fact, this globule harbors at its center a multiple protostellar system of young stellar objects (YSOs) at different stages of evolution. Near-IR observations towards the central IRAS source revealed the presence of two bright near-IR ($K$ band, 2.2 $\mu$m) objects classified as Class I protostellar candidates, \object{CB 54 YC1-I} (a confirmed Class I source; \citealt{Cia07})  and \object{CB 54 YC1-II}, plus a bright elongated feature (\object{CB 54 YC1-SW}) mainly seen in H$_{2}$[$v$=1-0 $S$(1)], 2.121 $\mu$m line \citep{Yun94,Yun95,Yun96,Kha03}. 
Water masers were detected by \cite{Gom06} and \cite{deG06} inside this southern elongated feature, suggesting the presence of an embedded protostar that pumps the maser emission.
This prediction was recently confirmed by the discovery of three faint mid-IR sources clustered near the position of the IRAS source and within the near-IR elongated feature \citep{Cia07}. They were named as \object{MIR-a}, \object{MIR-b}, and \object{MIR-c} and interpreted as very cool ($\simeq$ 100~K) Class 0 protostellar candidates of masses $\sim$1.5~$M_{\odot}$, $\sim$4~$M_{\odot}$, and $\sim$0.2~$M_{\odot}$ respectively. 

Water maser emission at 22 GHz is a good tracer of the mass-loss 
phenomena observed at the earliest stages of the formation of stars of all masses \citep{Rod80,Fel92,Xia95,Deb05}. In the case of low-mass objects, these water masers are usually associated with the youngest Class 0 protostars, produced by the interaction of powerful jets with a large amount of circumstellar material \citep{Fur01} and they tend to be located close to their powering source (within several hundred AU; \citealt{Che95,Cla98,Fur00,Fur03}). At those earliest stages of evolution young protostars show the most powerful molecular outflows \citep{Bon96}, which are believed to be driven by collimated jets \citep{Rag93}. The central objects that power the outflows are frequently
associated with weak and compact centimeter free-free continuum emission
from thermal radio jets \citep{Ang95,Ang96,Bel01}. These
radio jets trace the part of the outflow closest to the exciting source. These properties make the combination of water masers and radio continuum emission well suited for pinpointing the location of Class 0 protostars. 

In this work we present sensitive interferometric observations of water masers and radio continuum at 1.3 cm, using the Very Large Array (VLA). We also show radio continuum data at 3.6 and 6~cm from the VLA archive. The main goals of these observations were to  derive accurately the position of the water maser emission, to pinpoint the location of the exciting sources of the maser phenomenon, as well as to derive information about the driving engine of the molecular outflows that exist in the region.

This paper is structured as follows: in \S 2 we describe the observations
and data processing. In \S 3 we present and discuss the results derived from radio continuum and water masers observations. Finally, we present the conclusions of this work in \S 4.

\section{Observations and data processing}
Observations towards CB 54 were performed on 2005 January 22 and 31, and February 4 using the VLA of the National Radio Astronomy Observatory (NRAO)\footnote{The National Radio Astronomy Observatory is a facility of the National Science Foundation operated under cooperative agreement by Associated Universities, Inc.}  in the BnA configuration (project AG684). 
We observed simultaneously the
6$_{16}$-5$_{23}$ transition of H$_{2}$O (rest frequency = 22235.080
MHz) and continuum at 22285.080 MHz ($\simeq$~1.3 cm) using the
four IF spectral line mode and processing both right and left circular polarizations. 
For the H$_{2}$O observations we sampled 64 channels over a bandwidth of 3.125 MHz, centered at $V$$_{\rm LSR}$ = 15 km s$^{-1}$, with 0.66 km s$^{-1}$ velocity resolution. For continuum observations we used a bandwidth of 25 MHz that comprised 8 channels of 3.125 MHz. 
The total observing time including calibration was 4.5 hours per
day. The splitting of the observations into three different days was
required to reach the necessary sensitivity for the continuum data.
Our flux calibrator was 3C48, for which we adopted a flux density of
1.1 Jy using the latest VLA values (1999.2). The source J0609$-$157 was used as phase and bandpass calibrator (bootstrapped flux density = 3.90$\pm$0.08 Jy). 
The phase center of the observations was R.A.(J2000) =
07$^{h}$04$^{m}$21$\fs$4, Dec(J2000) = $-$16$\degr$23$\arcmin$15$\arcsec$.
The Astronomical Image and Processing System (AIPS), developed by
NRAO, was used to calibrate and process our data. We produced H$_{2}$O
line maps setting the ``robust'' weight parameter to 0, as a compromise
between angular resolution and sensitivity.
The size of the synthesized beam was $\simeq$0.25$\arcsec$$\times$0.14$\arcsec$ in the maps of each individual observing day. The water maser emission was strong enough to enable self-calibration. Spectral Hanning smoothing was applied to mitigate the Gibbs ringing,
which provided a final velocity resolution of 1.3 km s$^{-1}$. The
1.3 cm continuum data were cross calibrated using the self-calibration
solutions obtained from the line data for each individual day and were combined. Continuum maps were obtained using natural weighting to improve the signal to noise ratio, providing a synthesized beam of $\simeq$0.30$\arcsec$$\times$0.19$\arcsec$ (P.A. = 65$\rm{^{o}}$).

We have also reanalyzed radio continuum data at 8.44 GHz ($\simeq$~3.6 cm) and 4.86 GHz ($\simeq$~6 cm) from the VLA archive (these data were included in the papers by \citealt{Yun96} and \citealt{Mor97}).  
Both observations were performed with the array in the D configuration for projects AY071 and AY073.
A total bandwidth of 100 MHz was selected in the two sets of observations, and both right and left circular polarizations were processed. 
The time on source was $\simeq$ 20 minutes for 8.44 GHz data and $\simeq$ 1 hour for 4.86 GHz data. The source 3C48 was selected as the flux calibrator in both cases (adopted flux density equal to 3.2 Jy and 5.4 Jy respectively for 8.44 and 4.86 GHz frequencies). We have summarized the setup of these archived observations in Table~\ref{tbl-arc}.





\section{Results and discussion}
\subsection{Radio continuum emission}

\begin{figure}
\epsscale{0.4}
\rotatebox{-90}{
\plotone{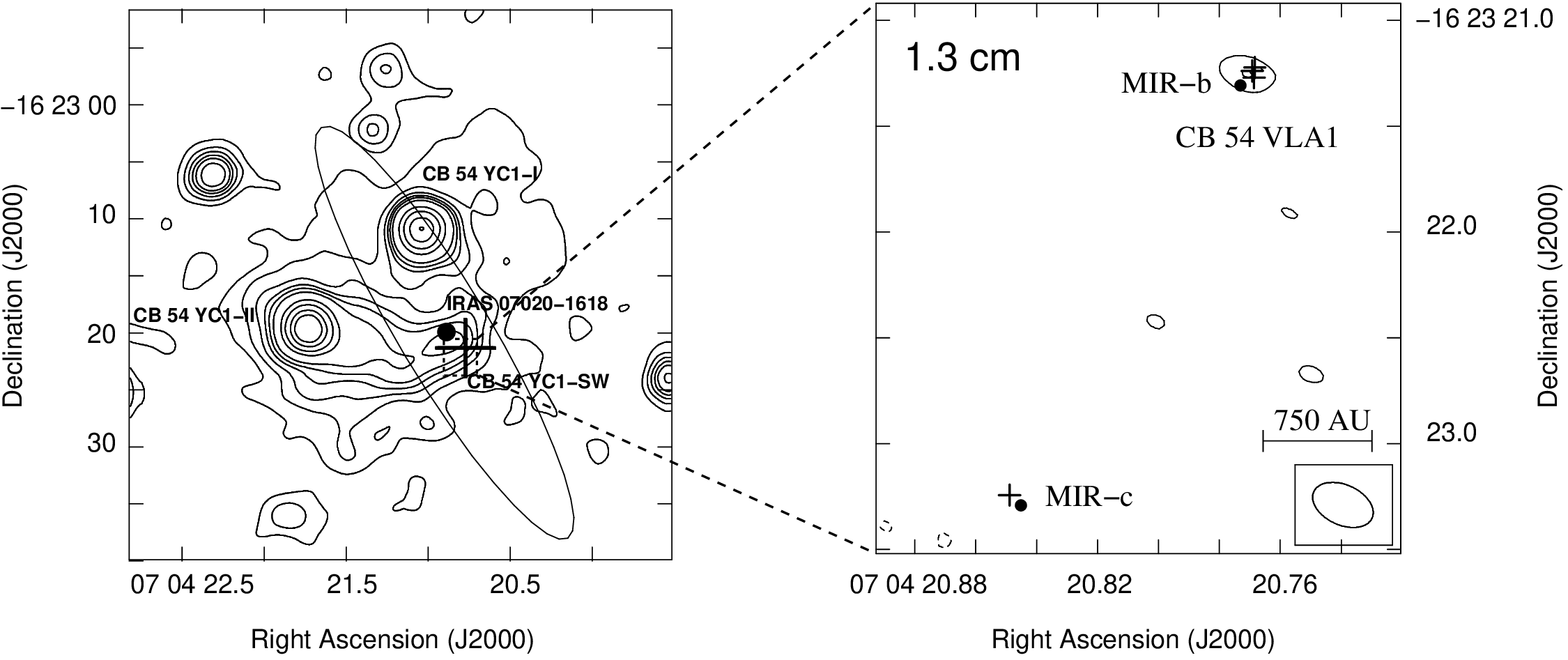}}
\caption{{\it Left}: Contour plot of the 2MASS near-IR emission towards CB~54, adapted from \cite{deG06}. The ellipse represents the positional error of IRAS~07020$-$1618, whose nominal position is marked with a filled circle. The cross represents the water maser emission detected by \cite{deG06}. {\it Right}: Contour map of the 1.3 cm continuum emission observed with the BnA configuration of the VLA. Levels are -3, 3, and 5 times 0.06 mJy beam$^{-1}$, the rms of the map. The HPBW of the synthesized beam is shown in the lower-right corner. Crosses mark the positions of the two groups of water masers detected in this work and observed simultaneously with the 1.3 cm continuum emission. Filled circles represent the mid-infrared sources MIR-b and MIR-c detected by \citet{Cia07}.} \label{1cm}
\end{figure}

We have detected a compact ($\leq$ 0.2$''$) continuum source at 1.3 cm (Table~$\ref{tbl-rc}$) at a position coinciding with the near-infrared elongated feature CB 54 YC1-SW (see Fig.~\ref{1cm}). This feature was proposed to trace an embedded YSO by \citet{deG06} on the basis of its water maser emission, a result that has been recently confirmed by the detection of three mid-infrared sources, within this feature, by \cite{Cia07}, who classified these objects as Class 0 protostellar candidates.  
Our 1.3~cm source, is spatially coincident with MIR-b (see Fig.~\ref{1cm}), one of the mid-infrared protostars detected by \cite{Cia07}.

\begin{figure}
\epsscale{0.5}
\plotone{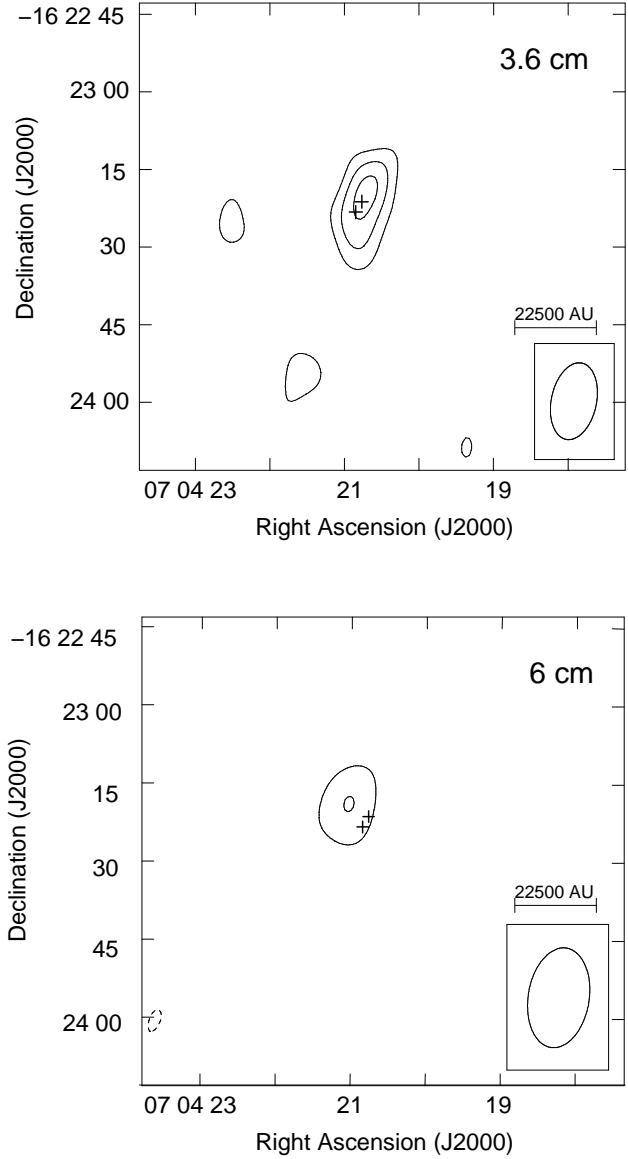}
\caption{Map of radio continuum emission observed at 3.6 and 6 cm. Contour levels are -3, 3, 5 and 7 times the rms of the map (0.025 mJy beam$^{-1}$) for data at 3.6 cm wavelength, and -3, 3 and 5  times the rms of the map (0.045 mJy beam$^{-1}$) for data at 6 cm. Crosses mark the positions of the two groups of water masers detected in this work. The HPBW of the synthesized beam is represented in the lower-right corner of each plot.} \label{3-6cm}
\end{figure}

Radio continuum emission at 3.6 and 6 cm is unresolved at both frequencies (see contour maps in Fig.~\ref{3-6cm}). In Table~$\ref{tbl-rc}$ we show detailed information about positions, flux densities, and uncertainties of the continuum emission presented in this section. From this analysis, we see that the position of the radio continuum emission at the three different frequencies is the same within the absolute positional errors, concluding that it comes from the same source, named as \object{CB 54 VLA1} by  \cite{Yun96} and \cite{Mor97}.

\subsubsection{Origin of the radio continuum emission}

In order to study the nature of the radio continuum emission associated with CB 54 VLA1, we compare the centimeter continuum luminosity inferred from the radio observations with the centimeter continuum luminosity expected from Lyman-continuum radiation from a ZAMS star of the given luminosity of the source. The bolometric luminosity derived from the flux densities of the source IRAS 07020$-$1618 close to our radio continuum source is $\sim$344 $L_{\odot}$ for an adopted distance of 1.5 kpc  \citep{Wan95}, which corresponds to a B5.5 ZAMS star \citep{Tho84}. We warn that the cluster of three YSOs detected by \cite{Cia07} falls within the positional error ellipsoid of IRAS 07020$-$1618 and they could contribute to the total luminosity of the IRAS source. Therefore we consider this value as an upper limit.  
The observed radio continuum luminosity at 1.3 cm wavelength (i.e., $S_{\nu}d^{2}$) is $\sim$ 7 $\times$ 10$^{-1}$  mJy kpc$^{2}$. On the other hand, assuming optically thin free-free emission from ionized hydrogen with an electron temperature of 10$^{4}$ K, we derive an expected upper limit of $S_{\nu}d^{2}$ $\lesssim$ 7 $\times$ 10$^{-3}$ mJy kpc$^{2}$ from a Lyman-continuum flux of $\sim$ 7 $\times$ 10$^{41}$ s$^{-1}$ (obtained from \citealt{Tho84} for a B5.5 ZAMS star). Thus, ionization by stellar photons fails by two orders of magnitude in explaining the observed radio emission, and another ionizing mechanism is required. This behavior has been observed before, for instance, by \cite{Tor85}, \cite{Rod89}, and  \cite{Ang95} for a large set of low-mass YSOs. 

A plausible mechanism for explaining the centimeter continuum emission
observed is the shock-ionization scenario proposed by \cite{Tor85}. In
this scenario the stellar wind responsible for a molecular outflow
generate shocks in the dense gas surrounding the central protostar and induces its ionization. \cite{Cur87, Cur89} modeled the shock-ionization scenario and derived the radio continuum emission under optically thin conditions. 
The spectral index measured by us in the 6$-$1.3 cm wavelength range is $\alpha$$_{6-1.3~\mbox{cm}}$=0.3$\pm$0.4 (where $S_\nu$ $\propto$ $\nu^{\alpha}$), consistent, within the errors, with optically thin free-free emission. The formulation of \cite{Cur87, Cur89} predicts a correlation between $S_{\nu}d^{2}$ and the momentum rate of the outflow {\it\.{P}}. Assuming that the  momentum rate in the outflow equals that in the stellar wind, {\it\.{P}}$=${\it\.{M}v} (where {\it\.{M}} is the mass loss rate of the wind, and $v$ the terminal velocity of the wind, adopting a typical value of 200 km s$^{-1}$), and a typical electron temperature in the ionized wind of 10$^{4}$ K, the prediction of the model gives:
\begin{equation}
 ~\dot P = \frac{10^{-3.5}}{\eta} ~S_{\nu}d^{2}    
\end{equation} 
with  $S_{\nu}d^{2}$ in mJy kpc$^{2}$ and $\dot P$ in M$_{\odot}$
yr$^{-1}$ km s$^{-1}$, being $S_\nu$ the flux density at 6 cm and
$\eta=\Omega/4\pi$ an efficiency factor that represents the fraction
of the stellar wind that is shocked and produces the observed radio
continuum emission. Scaling the outflow force derived by \cite{Yun94b}
from CO observations, to the adopted distance of 1.5 kpc, we obtain
$~\dot P$ = 4$\times$ 10$^{-4}$ $M_{\odot}$ km s$^{-1}$
yr$^{-1}$. Considering a radio continuum luminosity at 6 cm of 0.5 mJy
kpc$^{2}$, we derive and efficiency factor $\eta$$\simeq$0.4, which
indicates that the shock-ionization mechanism could explain the
observed radio continuum emission. We note that our estimate of
efficient factor $\eta$ can be affected by large errors, mainly due to
the uncertainty in the value of the  momentum rate of the outflow from molecular lines observations (see \citealt{Ang92} and \citealt{Ang95} for a detailed discussion of the dependence of the error in $\eta$ with the observational parameters).   
The efficiency factor derived in this work is somewhat higher than the
average value of $\eta$$\simeq$0.1 derived by \cite{Ang95} for a large
set of low luminosity objects. Nevertheless, the dispersion of the
efficiency values is relatively large and the best fit to that set of
data provides a value of $\eta$=10$^{-1\pm0.6}$ (adopting an
uncertainty of 2$\sigma$), i.e., $0.025 < \eta < 0.4$.


The observations presented here make CB 54 VLA1 a very good candidate
for driving a molecular outflow. Nevertheless, higher angular
resolution observations at centimeter and millimeter wavelengths
should be made to study the presence of a jet-disk system, since these
structures are typically observed with sizes of $\simeq$100 AU
($\simeq$ 0$\farcs$07 at a distance of 1.5 kpc), see \citet{Ang96}. In
particular, high angular resolution observations of 3.6 and 6 cm
emission, optically thicker than that at 1.3 cm, could better trace
low-brightness structures, and therefore, would be useful to prove the presence of a thermal radio jet with a morphology elongated in the same direction of the large scale molecular outflow. 
On the other hand, radio continuum millimeter data would be useful to reveal the presence of heated dust associated with a possible protoplanetary disk.

\subsection{Water maser emission}

\begin{figure}
\epsscale{1}
\plotone{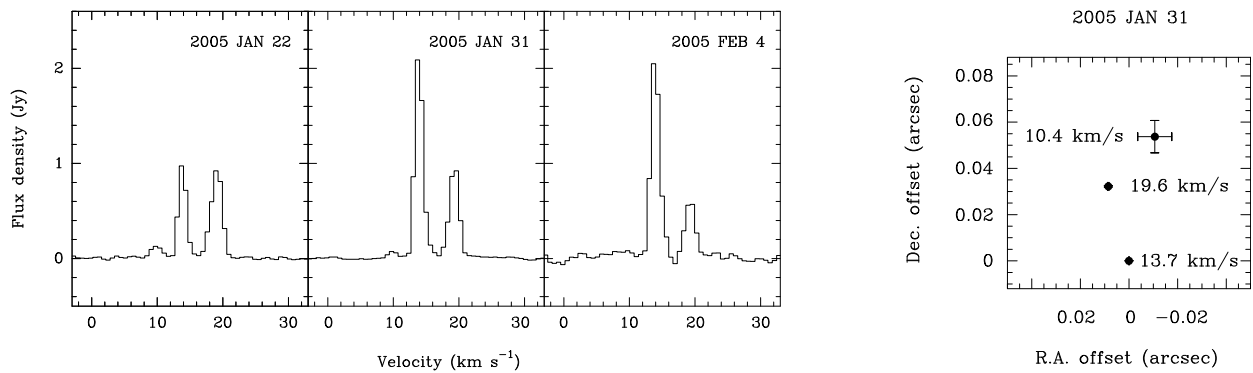}
\caption{{\it Left}: Spectra of the water maser emission observed with the VLA on 2005 January 22 (left), January 31 (center), and February 4 (right), integrated over a region of 2.1$\arcsec$$\times$2.7$\arcsec$, which includes the emission from the two groups of masers that are associated with MIR-b (CB54 VLA1) and MIR-c. {\it Right}: Positions of the independent water maser spectral features of the northern group of masers, which are associated with MIR-b (CB54 VLA1), as shown in the 1.3~cm map of Fig.~\ref{1cm}. The positions correspond to the 2005 January 31 observation. The (0,0) is the position of the reference feature used for self-calibration, and the size of the crosses represent the relative position uncertainty with respect to the reference feature (see Table 1). The southern group of water masers show one spot at 9.7 km~s$^{-1}$.} \label{spec-pos} 
\end{figure}

Table~\ref{tbl-masers} contains the results of our water masers observations with the  VLA. We observe three independent spectral features (see Fig.~\ref{spec-pos}, left panel). One of the features is approximately at the velocity of the cloud (V$_{\rm{LSR}}$=19.5 km s$^{-1}$; \citealt{Cle88}) and the rest are blue-shifted within 10 km s$^{-1}$ from the cloud velocity. All of them were detected on the three different days of observation. 
The masers are found at two different positions separated by
2.286$\arcsec$$\pm$0.004$\arcsec$ (distance from the southern component to the northern reference feature, $\simeq 3400$ AU at a distance of 1.5 kpc; see Fig.~\ref{1cm}). The northern group of masers is spatially associated with the mid-infrared object MIR-b and shows three spots at velocities 10.4, 13.7, and 19.6 km s$^{-1}$, separated by a few centi-arcseconds (see Fig.~\ref{spec-pos}, right panel). The southern group is composed by a single spot at a velocity of 9.7 km s$^{-1}$, spatially associated with the  mid-infrared object MIR-c.

The water maser emission in the region shows high variability, which
is typically observed in both low and high-mass young stellar objects
\citep{Rei81, Wil94, Cla96}. \citet{Gom06}, using the Robledo 70 m
antenna, detected a water maser spectrum composed of a single water
maser spectral feature observed at $V_{\rm LSR} = 13.7$ km s$^{-1}$ in
2002, at 7.9 km s$^{-1}$ in 2003, and at 8.7  km s$^{-1}$ in 2005. In addition,
\citet{deG06}, using the VLA, detected two different features at 15.8
and 17.8 km s$^{-1}$ in February 2004. In the observations reported
here, do not detect any of the maser spectral features observed in the
mentioned previous works in the region except the feature at 13.7 km
s$^{-1}$. This component shows a variation in its intensity by a
factor of $\sim$2 between 2005 January 22 and 2005 January 31 (see
Fig.~\ref{spec-pos}, left panel). On the other hand,  the features at
10.4 and 19.6 km s$^{-1}$ have not been reported before.  

Water masers associated with the northern YSO MIR-b are located at a distance $\leq$100 AU (assuming a distance of 1.5 kpc to the Bok globule) from the compact radio continuum source CB 54 VLA1 we detected at 1.3~cm, which suggests this object as the exciting source of the northern group of water masers. This short distance $\leq$100 AU between water masers and their exciting source is typically observed in a large set of low-mass star forming regions \citep{Che95, Cla98, Fur00}. The right panel of Fig.~\ref{spec-pos} shows the spatial distribution of the three northern spots obtained on the second day of observation, which corresponds to the data with the best signal to noise ratio. All the spots show similar positions on the three days of observation. They delineate a spatial structure of $\simeq$0.06$\arcsec$ (90 AU), elongated in the north-south direction. To ascertain whether the masers are associated with the molecular outflow or with disk material (i.e., whether they are tracing unbound or bound motions), for each maser component we estimate its velocity  ($V$) with respect to the mean velocity of the maser structure, 
and we calculate the mass ($M$) necessary to bind the gas that shows the maser
emission as $M = V^{2}RG^{-1}$, where $R$ is the distance of the maser structure to the central YSO, and we have conservatively assumed a value of $R\simeq100$ AU for the maser components (this distance to the center is an upper limit).  A mass of $M \simeq 2.8 M_{\sun}$ is enough to bind the gas responsible for the maser features. Since the mass of this source is estimated to be $\sim$ 4  $M_{\odot}$ \citep{Cia07}  we cannot discard bound motions.

On the other hand, the spatial association of the southernmost water maser emission with the mid-infrared YSO MIR-c suggests this object as the pumping source of the southern maser emission. In this case, we do not detect 1.3 cm continuum emission towards this object with a 3~$\sigma$ upper limit of 0.17 mJy.

\cite{Fur01} found that water masers in low-mass YSOs are usually excited by Class 0 sources due to the interaction of powerful jets with a large amount of circumstellar material.
Therefore, since sources that host water maser emission are good candidates for being in a very early stage of its evolution, as well as for being the exciting source of the mass-loss phenomenon, we propose the sources MIR-b (CB 54 VLA1) and MIR-c as the best candidates to be the driving engine of the molecular outflows that exist in this Bok globule.

\section{Conclusions}

We presented high angular resolution VLA observations of water masers and continuum emission at 1.3~cm towards the Bok globule CB 54. We complemented our observations with VLA archive data in the radio continuum at 3.6 and 6 cm. The main conclusions are the following: 

\begin{itemize}

\item  Our subarcsecond angular resolution observations at 1.3 cm allow us to establish that the radio continuum emission detected to the south-west of the Bok globule is associated with the mid-infrared source MIR-b. The spectral index of the emission between 6 and 1.3 cm is flat, consistent with optically thin free-free emission from ionized gas. A shock scenario mechanism is needed to produce the radio continuum luminosity at cm wavelengths.

\item Water masers are found in two different regions. The northern group of masers coincides within $<$ 100 AU with the source  CB~54 VLA1, which is associated
with the mid-IR protostar MIR-b, and whose position at 1.3 cm is reported in this paper. The southern region of water maser emission is located $\sim2.3''$ SW, towards the position of the faint mid-IR 
object MIR-c, without detectable radio continuum emission.

\item The association of the mid-IR sources MIR-b and MIR-c with water masers, confirms the embedded protostellar nature of both objects and suggests these protostars are the best candidates for driving the molecular outflows observed in the region.
  
\end{itemize}

\acknowledgments

We thank the referee for providing constructive comments and help in improving the contents of this paper. We are thankful to Per Bergman for his suggestions. GA, IdG, JFG, JMT, and OS are partially supported by Ministerio de Ciencia e Innovaci\'on (Spain), grant AYA 2008-06189-C03 (including FEDER funds), and by Consejer\'{\i}a de Innovaci\'on, Ciencia y Empresa of Junta de Andaluc\'{\i}a, (Spain). The work by TBHK was performed in part at the Jet Propulsion Laboratory under contract between the National Aeronautics and Space Administration and the California Institute of Technology.

{\it Facilities:} \facility{VLA ()}



\clearpage

\begin{deluxetable}{cccrccccc}
\tabletypesize{\scriptsize}
\tablecaption{Setup of VLA archive data analyzed\label{tbl-arc}}
\tablewidth{0pt}
\tablehead{
\colhead{Freq.}&
\colhead{Observation}   &
\colhead{Project}   &
\colhead{R.A.\tablenotemark{a}}&
\colhead{Dec.\tablenotemark{a}}&
\colhead{Beam Size} &
\colhead{P.A.}   &
\colhead{Phase}& 
\colhead{$S_{\rm cal}$\tablenotemark{b}}

\\
\colhead{(GHz)}&  
\colhead{date}   &
\colhead{name}   &
\colhead{(J2000)}&
\colhead{(J2000)}&
\colhead{($\arcsec$$\times$$\arcsec$)} &
&  
\colhead{calibrator}&
\colhead{(Jy)} 

}
\startdata
8.44    &19-MAY-1995  &AY071  &07$^{h}$04$^{m}$20$\fs$9  &$-$16$\degr$23$\arcmin$20$\arcsec$  &15$\times$9  &$-$12$\degr$   &J0609$-$157   &5.60$\pm$0.01   \\
4.86    &30-JUN-1996  &AY073  &07$^{h}$04$^{m}$21$\fs$2  &$-$16$\degr$23$\arcmin$15$\arcsec$  &19$\times$12 &$-$9$\degr$   &J0729$-$366   &2.845$\pm$0.006   \\

\enddata
\tablenotetext{a}{Coordinates of the phase center.}
\tablenotetext{b}{Bootstrapped flux densities of phase calibrators.}

\end{deluxetable}

\clearpage

\begin{deluxetable}{cllcc}
\tablecaption{Radio continuum emission towards CB~54\label{tbl-rc}}
\tablewidth{0pt}
\tablehead{
\colhead{Frequency}&
\colhead{R.A.}&
\colhead{Dec.}&
\colhead{Position\tablenotemark{a}}&
\colhead{Flux density} 
\\
\colhead{(GHz)}&  
\colhead{(J2000)}&
\colhead{(J2000)}&
\colhead{uncertainty ($''$)}&
\colhead{(mJy)} 
}
\startdata
22.24    &07$^{h}$04$^{m}$20$\fs$771  &$-$16$\degr$23$\arcmin$21$\farcs$26  &0.08   &0.29$\pm$0.11\\
8.44     &07$^{h}$04$^{m}$20$\fs$73  &$-$16$\degr$23$\arcmin$21$\farcs$7    &2.3    &0.19$\pm$0.05\\
4.86     &07$^{h}$04$^{m}$21$\fs$0  &$-$16$\degr$23$\arcmin$19$\arcsec$              &4      &0.23$\pm$0.09\\
\enddata

\tablecomments{Uncertainties are 2$\sigma$.}
\tablenotetext{a}{Absolute position error.} 
\end{deluxetable}

\clearpage

\begin{deluxetable}{lcccrr}
\tabletypesize{\small}
\tablecaption{Water Maser Features Detected with the VLA towards CB 54\label{tbl-masers}}
\tablewidth{0pt}
\tablehead{
\colhead{Date}&
\colhead{R.A. Offset\tablenotemark{a}} &
\colhead{Dec.Offset \tablenotemark{a}}& 
\colhead{Position\tablenotemark{b}} &
\colhead{Flux Density}   &
\colhead{$V_{\rm LSR}$\tablenotemark{c}}
\\
 &  
\colhead{($\arcsec$)}&
\colhead{($\arcsec$)}&\colhead{Uncertainty ($\arcsec$)} &\colhead{(mJy)}& 
\colhead{(km s$^{-1}$)} 

}

\startdata
2005 JAN 22   & 0.0094      &0.0318   &0.0004                      &913$\pm$3   &19.6\\  
              & 0               &0    &\nodata\tablenotemark{d}    &928$\pm$3   &13.7\\  
              & -0.011      &0.046    &0.012                       &31$\pm$3   &10.4\\        
              & 1.152       &-1.972   &0.004                       &106$\pm$3   &9.7 \\

\tableline
& & & & &\\

2005 JAN 31   &0.00848    &0.03220    &0.00027   &913$\pm$2  &19.6\\ 
              &0          &0    &\nodata\tablenotemark{d}   &2073$\pm$2    &13.7\\ 
              &-0.011      &0.057      &0.007     &35$\pm$2  &10.4\\                          
              &1.152      &-1.974     &0.003     &74$\pm$2  & 9.7\\

\tableline
& & & & &\\

2005 FEB 04   &0.0093     &0.0328  &0.0013    &562$\pm$6    &19.6\\    
              &0          &0       &\nodata\tablenotemark{d}    &2080$\pm$6    &13.7\\    
              &-0.014      &0.056   &0.023     &33$\pm$6  &10.4\\    
              &1.149      &-1.971  &0.013     &57$\pm$6  &9.7   \\

\enddata
\tablecomments{Quoted uncertainties in this table are $2\sigma$}.
\tablenotetext{a}{Right Ascension and Declination offsets of the peak of each distinct water maser spectral feature with respect to the reference feature used for self-calibration, whose position is R.A.(J2000.0)=07$^{h}$04$^{m}$20$\fs$769, Dec.(J2000.0)=$-$16$\degr$23$\arcmin$21$\farcs$27, in the three observation days. The absolute position uncertainty of this reference position is $\simeq$ 0$\farcs$05.}
\tablenotetext{b}{Uncertainty in the relative positions with respect to the reference position.}
\tablenotetext{c}{{\rm L}SR velocity of the spectral features. Velocity resolution is $\sim$ 1.3 km s$^{-1}$.}
\tablenotetext{d}{Reference feature.}


\end{deluxetable}

\end{document}